\documentstyle[12pt,equation]{article}
\begin{document}
\newcommand{\tcr}{T_{cr}}
\newcommand{\bb}{\bar{B}}
\newcommand{\cb}{\bar{C}}
\newcommand{\mb}{\bar{\mu}^2}
\newcommand{\lb}{\bar{\lambda}}
\newcommand{\rrz}{\rho_0}
\newcommand{\rrzk}{\rho_0(k)}
\newcommand{\rrzkt}{\rho_0(k,T)}
\newcommand{\rrzt}{\rho_0(T)}
\newcommand{\ct}{\tilde{c}}
\newcommand{\be}{\begin{equation}}
\newcommand{\ee}{\end{equation}}
\newcommand{\een}{\end{subequations}}
\newcommand{\ben}{\begin{subequations}}
\newcommand{\beq}{\begin{eqalignno}}
\newcommand{\eeq}{\end{eqalignno}}
\renewcommand{\thefootnote}{\fnsymbol{footnote} }
\pagestyle{empty}
\noindent
DESY 93-004 \\
January 1993
\vspace{3cm}
\begin{center}
{{ \Large  \bf
The Large $N$ Limit \\
and the High Temperature Phase Transition \\
for the $\phi^4$ Theory
}}\\
\vspace{10mm}
M. Reuter, N. Tetradis \\
{\em Deutsches Elektronen-Synchrotron DESY, Gruppe Theorie,\\
Notkestr. 85, D-2000 Hamburg 52} \\
\vspace{5mm}
and \\
\vspace{5mm}
C. Wetterich \\
{\em
Institut f\"ur Theoretische Physik, Universit\"at Heidelberg,\\
Philosophenweg 16, D-6900 Heidelberg}\\
\end{center}
\setlength{\baselineskip}{20pt}
\setlength{\textwidth}{13cm}

\vspace{3.cm}
\begin{abstract}

{
We study,
with various methods,
the high temperature phase transition for the
$N$-component $\phi^4$ theory in the large $N$ limit.
Our results fully confirm a previous investigation of the
problem, for arbitrary $N$, with the method of the average potential.
The phase transition is of the second order with an effectively
three-dimensional critical behaviour.
}
\end{abstract}
\clearpage
\setlength{\baselineskip}{15pt}
\setlength{\textwidth}{16cm}
\pagestyle{plain}
\setcounter{page}{1}

\newpage

\setcounter{equation}{0}
\renewcommand{\theequation}{{\bf 1.}\arabic{equation}}

\section*{1. Introduction}

\setcounter{equation}{0}
\renewcommand{\theequation}{{\bf 2.}\arabic{equation}}

The $N$-component $\phi^4$ theory is the prototype for investigations
concerning the question of symmetry restoration at high temperature.
Following the original argument of Kirzhnits and Linde
on the restoration of gauge symmetry \cite{linde1},
it was considered in
all subsequent investigations of the problem \cite{linde2,weinberg,doljak}.
The physical importance of the theory
arises from the fact that, for $N=4$,
it describes
the scalar
sector of the electroweak standard model
in the limit of vanishing gauge and Yukawa couplings, and
is relevant for the QCD phase transition
for vanishing up and down quark mass \cite{wilczek}.
For a negative bare mass term the theory can exhibit spontaneous breaking
of the $O(N)$ symmetry to $O(N-1)$. It was demonstrated in
\cite{linde2}-\cite{weinberg} that the original symmetry is restored
at sufficiently high temperature. This was achieved
through the
perturbative
calculation of the effective potential \cite{colwein},
and its
generalization for non-zero temperature.
However, it was realized in the original works
that the perturbative expansion breaks down near the critical temperature.
This is due to the presence of infrared divergencies which
destroy the convergence of the perturbative series.
As a result the details of the phase transition cannot be discussed
in any finite order in perturbation theory, and even the order
of the transition cannot be predicted.
An amelioration of the
situation was achieved in \cite{doljak} with the summation of an
infinite subclass of diagrams (the ``daisy'' graphs) in the large $N$ limit,
but for small $N$ the question remained open.
Different perturbative calculations led to predictions of a first
order phase transition \cite{wrong}, while improved treatments
reconfirmed the breakdown of the perturbative expansion near the
critical temperature \cite{zwirner}.
The situation was eventually resolved in \cite{transition}
through the method of the average potential
which employs renormalization group
ideas \cite{christof,convex,fermiongauge}.
This method leads to an evolution equation for the dependence
of the effective potential on a variable infrared cutoff, which
provides a good handling of the
problematic infrared regime by effectively resumming all the relevant
classes of diagrams of perturbation theory.
The high temperature phase transition
for the four-dimensional $\phi^4$ theory was shown to be of the second order
belonging to the same universality class as the
phase transition of the three-dimensional
$O(N)$-symmetric $\sigma$-models.
A detailed quantitative picture was obtained, including the
critical exponents for the behaviour in the vicinity of the
critical temperature.
\par
The main purpose of this work is to confirm the results of
\cite{transition} through the use of alternative
methods. As it was explained before, such methods
should go beyond any finite order in perturbation theory.
It is for this reason that the large $N$ limit is particularly useful.
In this limit the functional integral which defines the effective
potential can be explicitly evaluated. In an alternative approach the
Schwinger-Dyson equations of the theory can be solved
in the critical region.
In this way we shall be able to obtain a completely satisfactory picture
of the phase transition in order to verify the results of
\cite{transition}.
We shall compare with
the work of ref.
\cite{jain}, where the large $N$
limit of the $\phi^4$ theory has also been discussed.
Going beyond the leading $1/N$ result is more difficult and
the use of the renormalization group becomes indispensable.
We shall demonstrate the power of the technique
employed in \cite{transition} by using it in order to calculate the critical
exponents for small $N$.
\par
The paper is organized as follows: In section 2 we first summarize the
formalism of the average potential for the $N$-component $\phi^4$
theory. Then we use standard large $N$ techniques in order to
obtain a set of equations which permit a straightforward calculation of
the average potential. As a result we show how the average potential
interpolates between the classical and the effective potential.
Subsequently we obtain
an exact
renormalization group equation for the average potential, which can
be used in order to solve for the effective potential
with the
classical potential as a boundary condition. This equation
is identical to the one employed in \cite{transition}
in the large $N$ limit.
We also generalize our formalism to the case of
non-zero temperature.
In section 3 we
compute directly the large $N$ limit of
the
effective potential in the critical region (temperatures close to the
critical one) without using the concept of the average potential.
In section 4 an alternative approach
is followed, which solves the Schwinger-Dyson equations
in order to
reproduce the results of section 3.
Sections 3 and 4 are self-contained and can be read independently of the
rest of the paper.
In section 5
we demonstrate the exact agreement
of the results of the previous two sections with \cite{transition}.
Then we make use of the renormalization group equation
in order to go
beyond the leading order in $1/N$ and obtain the critical exponents
for small $N$.
Our conclusions are given in section 6.

\section*{2. The average potential in the large $N$ limit}

We consider a theory of $N$ real scalar fields $\chi^{a},
a=1...N$, with an $O(N)$-invariant action of the form
\be
S[\chi] = \int d^d x
\bigl\{ \frac{1}{2} \partial_{\mu} \chi^a \partial^{\mu} \chi_a
+ V(\xi) \bigr\},~~~~~~~~~~~~~ \xi = \frac{1}{2} \chi^a \chi_a.
\label{twoone} \ee
For the time being the dimensionality of the Euclidean space-time,
$d$, is kept arbitrary.
Following ref. \cite{christof},
we obtain from the above fundamental (or
``microscopic'') action $S$ an average (or ``block spin'') action $\Gamma_k$
\be
\exp\{-\Gamma_k[\phi]\} = \int {\cal D} \chi
P_k[\phi,\chi] \exp\{-S[\chi]\}.
\label{twotwo}  \ee
The constraint operator
$P_k[\phi,\chi]$ enforces the average of $\chi$
over a volume with size $\sim k^{-d}$
to equal the average field $\phi$ up to small fluctuations.
It is given by the expression
\be
P_k[\phi,\chi] = \exp \bigl\{ - \frac{1}{2} \int
d^d x [\phi^a - f_k(-\Box) \chi^a] P(-\Box) [\phi_a - f_k(-\Box) \chi_a]
\bigr\}.
\label{twothree} \ee
Here $f_k(q^2)$ is the
Fourier transform of the function used in order to implement the averaging
of $\chi$ and
different ``averaging schemes'' correspond to different choices of $f_k$.
We consider the two parameter family of schemes
\be
f_k(q^2) = \exp \bigl\{ - a \left( \frac{q^2}{k^2} \right)^{b} \bigr\},
\label{twofour} \ee
(with $a \sim 1$ and $b > 2$ \cite{christof}).
$P$ is given by
\be
P(q^2) = q^2 [1 - f^2_k(q^2)]^{-1},
\label{twofive} \ee
and has a minimum
\be
{\rm min}P(q^2) \sim k^2.
\label{twofivea} \ee
In eq. (\ref{twothree})
$q^2$ is replaced by $- \Box$ and the resulting operator acts on
$\chi^a(x)$ in position space. The right-hand-side of eq. (\ref{twothree})
contributes a factor $q^2 f_k^2 (1-f^2_k)^{-1}$ to the term quadratic in
$\chi^a$. Combining this term with the kinetic term from the classical action
(\ref{twoone}) we obtain
\beq
\exp\{ - \Gamma_k[\phi] \} = &\exp \bigl\{ - \frac{1}{2} \int d^d x
\phi^a  P(- \Box) \phi_a  \bigr\} \nonumber \\
&\cdot \int {\cal D}\chi \exp \biggl\{- \int d^d x \left(
\frac{1}{2} \chi^a P(- \Box) \chi_a -
\chi^a f_k(- \Box) P(- \Box) \phi_a + V(\xi) \right)
\biggr\}. \nonumber \\
{}~&~~
\label{twosix} \eeq
\par
In order to implement the large $N$ approximation we introduce two
(scalar, $O(N)$-singlet) auxiliary fields $B(x)$ and $C(x)$
and write \cite{largen1}
\be
\exp \bigl\{ - \int d^dx V(\xi) \bigr\} =
\int {\cal D}B {\cal D}C \exp \biggr\{ - \int d^dx \left( C(\xi-B) + V(B)
\right) \biggl\}.
\label{twoseven} \ee
Here we have used a rotated integration contour for the delta-functional
\be
\delta[X] = \int {\cal D}C \exp\bigr\{ - \int d^d x C X \bigl\}.
\label{twoeight} \ee
By inserting (\ref{twoseven}) into (\ref{twosix}) the action becomes
quadratic in $\chi$ and the integration can be performed easily.
We obtain
\be
\exp\{ - \Gamma_k[\phi] \} =
\int {\cal D}B {\cal D}C \exp \bigr\{
- S_{aux}[B,C;\phi]
\bigl\},
\label{twonine} \ee
with
\beq
S_{aux}&[B,C;\phi] = \nonumber \\
&\int d^d x \biggl\{ - \frac{1}{2} \phi^a \left(
f_k(-\Box)P(-\Box) \frac{1}{P(-\Box)+C} f_k(-\Box)P(-\Box) - P(-\Box) \right)
\phi_a
+ V(B) - B C \biggr\}
\nonumber \\
&+ \frac{N}{2} \ln {\rm det} \left[ P(-\Box) + C \right].
\label{twoten} \eeq
The last term of eq. (\ref{twoten}) originates in the internal
$\chi$-loops and is proportional to $N$.
The other terms in (\ref{twoten})
are not manifestly proportional to $N$, but this can
be achieved by assuming an appropriate $N$ dependence
for the fields:
$\phi \sim \sqrt{N}$, $B \sim N$
\footnote{
Formally one has to rescale $B \rightarrow B/N$ and consider
$S_{aux}[B,C;\sqrt{N}\phi]$ instead of $S_{aux}[B,C;\phi]$.
Our naive assumption for the $N$ dependence of $\phi$ and $B$
reproduces the effects of this procedure, while simplifying the
counting of the $N$ dependence of the various terms.
}.
For a polynomial
potential of the form
\be
V(B) = \sum_p g_{2p} B^p,
\label{twoeleven} \ee
the contribution of the potential is $\sim N$
if the generalized couplings $g_{2p}$
have an $N$ dependence
\be
g_{2p} \sim N^{1-p}.
\label{twotwelve} \ee
More specifically, the quartic coupling $g_4 \equiv \lambda/2$ must scale as
$g_4 \sim 1/N$ for large $N$ \cite{largen2}.
\par
Since now $S_{aux} \sim N$, in the limit $N \rightarrow \infty$ the functional
integral (\ref{twoten}) is dominated by its stationary point. Therefore
\be
\Gamma_k[\phi] = S_{aux} \left[\bb(\phi),\cb(\phi);\phi \right],
\label{twothirteen} \ee
where
$\bb$ and $\cb$ are the $\phi$-dependent solutions of the saddle point
conditions
\beq
\frac{\delta S_{aux}}{\delta B}[\bb,\cb] = &0,
\label{twofourteena} \\
\frac{\delta S_{aux}}{\delta C}[\bb,\cb] = &0.
\label{twofourteenb} \eeq
We are interested in
translationally invariant states of the theory which correspond to
constant $\phi$, $\bb$, $\cb$.
For such states the average action $\Gamma_k$
is related to the average potential $U_k$ through the expression
\footnote{
The wavefunction renormalization is zero in the
large $N$ limit.  For this reason we
do not explicitly discuss the derivative terms in the effective action.
}
\be
\Gamma_k[\phi] = \int d^dx U_k(\rho).
\label{twofifteen} \ee
Here $\rho$
is an abbreviation which we shall frequently use in the following
\be
\rho = \frac{1}{2} \phi^a \phi_a.
\label{twosixteen} \ee
For constant $\phi$ and $\cb$
the first term in eq. (\ref{twoten}) simplifies
considerably since
\be
-\frac{1}{2} \phi^a \left[f_k P \frac{1}{P+\cb} f_k P - P \right] \phi^a
= \rho \cb.
\label{twoseventeen} \ee
In order to derive the above expression we have exploited that $P(q^2)$
diverges
$P(q^2) \sim \left( q^2 \right)^{1-b}$ for $q^2 \rightarrow 0$.
We also notice that
for constant $\cb$ the operator $P(- \Box ) + \cb$ is diagonal in momentum
space.
Combining all of the above we find
\be
U_k(\bb,\cb;\rho) = V(\bb) + (\rho-\bb)\cb + \frac{N}{2}
\int_{\Lambda} \frac{d^dq}{(2 \pi)^d} \ln \left[ P(q^2) + \cb \right].
\label{twoeighteen} \ee
An ultraviolet cutoff $\Lambda$ is implied for the momentum integration in the
previous expression, and $\bb,\cb$ are assumed to be solutions of the
saddle point conditions (\ref{twofourteena}), (\ref{twofourteenb}), which read
\beq
\cb = &V'(\bb), \label{twonineteen} \\
\bb = &\rho + \frac{N}{2}
\int_{\Lambda} \frac{d^dq}{(2 \pi)^d} \frac{1}{ P(q^2) + V'(\bb)}.
\label{twotwenty} \eeq
The last equation gives $\bb$ as a function of $\rho$.
Eliminating $\cb$ from the expression for $U_k$, we finally obtain
for the average potential in the large $N$ limit
\be
U_k(\rho) = V(\bb) + (\rho-\bb) V'(\bb) + \frac{N}{2}
\int_{\Lambda} \frac{d^dq}{(2 \pi)^d} \ln \left[ P(q^2) + V'(\bb) \right],
\label{twotwentyone} \ee
with $\bb(\rho)$ determined through the ``gap equation'' (\ref{twotwenty}).
For $k=\Lambda$ eqs. (\ref{twotwenty}), (\ref{twotwentyone})
assume the simple form
\beq
\bb = &\rho, \nonumber \\
U_{\Lambda}(\rho) = &V(\rho).
\label{twotwentyonea} \eeq
In this limit the average potential coincides with the classical potential.
For $k \rightarrow 0$ the modified inverse propagator $P(q^2)$ of eqs.
(\ref{twofour}), (\ref{twofive}) approaches the standard form
$P(q^2)=q^2$. In the same time the average potential
$U_k$ approaches the convex effective potential $U$ \cite{christof,convex}.
In this limit eqs. (\ref{twotwenty}), (\ref{twotwentyone})
coincide with the well known
\cite{largen1} results for the effective potential in the leading order of
the $1/N$ expansion.
An interesting formula holds for
the $\rho$-derivative of $U_k$. Differentiating
(\ref{twotwentyone})
and making use of (\ref{twotwenty}) we obtain
\be
\frac{d}{d \rho} U_k(\rho) \equiv U'_k(\rho) = V' \left( \bb(\rho) \right).
\label{twotwentysix} \ee
\par
We now proceed to derive a renormalization group equation for the
average potential $U_k$, which is exact in the limit
$N \rightarrow \infty$. We shall see that this equation coincides with
the one obtained through the ``renormalization group improved''
one-loop calculation \cite{christof,convex,transition}.
We start by investigating how $U_k$ responds to a change of the scale
$k$. Taking the logarithmic derivative with respect to $k$ of
eq. (\ref{twotwentyone})
and making use of the ``gap equation'' (\ref{twotwenty})
we find
\be
k \frac{d}{dk} U_k(\rho) =
\frac{N}{2}
\int_{\Lambda} \frac{d^dq}{(2 \pi)^d} \frac{1}{ P(q^2) + V'(\bb)}
k \frac{d}{dk} P(q^2).
\label{twotwentyfour} \ee
Combining eqs. (\ref{twotwentysix}), (\ref{twotwentyfour}) leads to the
following integro-differential equation
\be
k \frac{d}{dk} U_k(\rho) =
\frac{N}{2}
\int_{\Lambda} \frac{d^dq}{(2 \pi)^d} \frac{1}{ P(q^2) + U'_k(\rho)}
k \frac{d}{dk} P(q^2) =
N v_d
\int_{0}^{\Lambda^2} {dx} \frac{x^{\frac{d}{2}-1}}{ P(x) + U'_k(\rho)}
k \frac{d}{dk} P(x),
\label{twotwentyseven} \ee
where
\be
v_d = \left[ 2^{d+1} \pi^{\frac{d}{2}} \Gamma \left( \frac{d}{2} \right)
\right]^{-1}.
\label{twotwentyfive} \ee
This is the desired exact evolution equation
for $U_k$.
It describes how the average potential
evolves as the infrared cutoff is lowered from
$k=\Lambda$ to $k=0$ and successive modes are integrated out.
With the initial condition
$U_{k=\Lambda}(\rho) = V(\rho)$, it provides the means for calculating
the effective potential $U(\rho) = U_{k=0}(\rho)$.
The form of the inverse propagator
$P(q^2)$, given by eqs. (\ref{twofour}), (\ref{twofive}), (\ref{twofivea}),
provides for an effective infrared cutoff for all the $\chi$-modes
with $q^2 < k^2$.
Also for modes with $q^2 > k^2$ the inverse propagator is
essentially independent of $k$.
As a result, only the modes with
$q^2 \sim k^2$ contribute to the
integral in (\ref{twotwentyseven}).
Eq. (\ref{twotwentyseven}) can be compared to the evolution
equation
obtained through the ``renormalization group improved''
one-loop calculation and used for
investigations of the average potential in
\cite{christof,convex,transition}.
The two equations are identical in the limit $N \rightarrow \infty$.
\par
We finally consider a specific
classical potential
$V(\xi)$
\be
V(\xi) = - \mb \xi + \frac{1}{2} \lb \xi^2,~~~~~~~~~~~~
\xi = \frac{1}{2} \chi^a \chi_a,~~\mb>0.
\label{twothirty} \ee
For the above form of the classical
potential
(\ref{twotwenty}), (\ref{twotwentyone}), (\ref{twotwentysix})
result in the following equation for the average potential
\be
U'_k(\rho) = -\mb + \lb \rho + \frac{N}{2} \lb
\int_{\Lambda} \frac{d^d q}{(2 \pi)^d} \frac{1}{ P(q^2) + U'_k(\rho)}.
\label{twothirtyone} \ee
For $k=0$, where $P(q^2)=q^2$, this is the well known
result for the effective potential in leading order in the
$1/N$ expansion \cite{largen1}. Here it is generalized to the average potential
for arbitrary values of the scale $k$.
By performing a first mass renormalization
\be
\mu_1^2 = \mb
- \frac{N}{2} \lb
\int_{\Lambda} \frac{d^d q}{(2 \pi)^d} \frac{1}{q^2}
\label{twothirtytwo} \ee
we find
\be
U'_k(\rho) = -\mu^2_1 + \lb \rho
+ \frac{N}{2} \lb
\int_{\Lambda} \frac{d^d q}{(2 \pi)^d} \frac{q^2-P(q^2)}{q^2 P(q^2)}
- \frac{N}{2} \lb U'_k(\rho)
\int_{\Lambda} \frac{d^d q}{(2 \pi)^d} \frac{1}{P(q^2)
\left(P(q^2) + U'_k(\rho) \right) }.
\label{twothirtythree} \ee
The integrals in the above expression are ultraviolet and infrared
finite in three dimensions. We can, therefore, take
$\Lambda \rightarrow \infty$ in (\ref{twothirtythree}).
(In four dimensions additional logarithmic renormalizations
must be performed.) Evaluation of the integrals gives an implicit
equation for $U'_k(\rho)$ as function of $\rho$, $k$, $\mu^2_1$ and
$\lb$. This can be used as a check of results obtained through different
approaches in the limit $N \rightarrow \infty$.
In particular, we find for the $k$ dependence of the quartic coupling
\be
U''_k(\rho) = \lb \left( 1 +
\frac{N}{2} \lb
\int_{\Lambda} \frac{d^d q}{(2 \pi)^d} \frac{1}
{\left( P(q^2)+U'_k(\rho) \right)^2 }
\right)^{-1}.
\label{twothirtyfour} \ee
\par
In order to extend our discussion to the non-zero temperature problem
we only need to recall that, in Euclidean formalism, non-zero temperature $T$
results
in periodic boundary conditions in the time direction (for bosonic fields),
with periodicity $1/T$ \cite{kapusta}.
This leads to a discrete spectrum for the zero component of the momentum
$q_0$
\be
q_0 \rightarrow 2 \pi m T,~~~~~~~~~m=0,\pm1,\pm2,...
\label{thrthree} \ee
As a consequence the integration over $q_0$ is replaced by summation over the
discrete spectrum
\be
\int \frac{d^d q}{(2 \pi)^d} \rightarrow
T \sum_m \int \frac{d^{d-1}\vec{q}}{(2 \pi)^{d-1}}.
\label{thrfour} \ee
With the above remarks in mind we can easily generalize our
results
in order to take into account the temperature effects.
Eq. (\ref{twothirtyone}) is replaced by
\be
U'_k(\rho,T) = -\mb + \lb \rho + \frac{N}{2} \lb
T \sum_m \int_{\Lambda} \frac{d^{d-1} \vec{q}}{(2 \pi)^{d-1}}
\frac{1}{ P(\vec{q}^2 + 4 \pi^2 m^2 T^2) + U'_k(\rho,T)}
\label{thrfffive} \ee
in an obvious notation. (The cutoff $\Lambda$ is taken
$\Lambda \gg T$.)
Again, the usual temperature dependent effective potential
\cite{linde2}-\cite{weinberg}
obtains from $U_k(\rho,T)$ in the limit $k \rightarrow 0$.
The generalization of the evolution equation (\ref{twotwentyseven})
reads
\be
k \frac{d}{dk} U_k(\rho,T) =
\frac{N}{2} T \sum_m
\int_{\Lambda} \frac{d^{d-1} \vec{q}}{(2 \pi)^{d-1}}
\frac{1}{ P(\vec{q}^2 + 4 \pi^2 m^2 T^2) + U'_k(\rho,T)}
k \frac{d}{dk} P(\vec{q}^2 + 4 \pi^2 m^2 T^2).
\label{fivlalatwo} \ee
\par
The results of this section can be summarized in eqs.
(\ref{twotwenty}), (\ref{twotwentyone}) and (\ref{twotwentyseven}).
The first two determine the average potential $U_k(\rho)$ for any value of
the scale $k$ in the large $N$ limit. The third one provides an
exact equation for the evolution of the average potential
as a function of $k$ in the same limit.
We also point out the simple relation between the
derivative of the average potential and the classical one
(\ref{twotwentysix}).
In the specific case of the classical potential (\ref{twothirty}),
eqs. (\ref{twotwenty}), (\ref{twotwentyone})
are replaced by (\ref{twothirtyone}).
For non-zero temperature
eqs. (\ref{twothirtyone}), (\ref{twotwentyseven})
take the form
(\ref{thrfffive}), (\ref{fivlalatwo}) respectively.
In the following sections we shall use these results in order to study the
high temperature phase transition for the
four-dimensional, $O(N)$-symmetric $\phi^4$ theory.

\setcounter{equation}{0}
\renewcommand{\theequation}{{\bf 3.}\arabic{equation}}

\section*{3. The high temperature phase transition
in the large $N$ limit}

{}From this point on we restrict our discussion to four dimensions
and to the classical potential (\ref{twothirty}).
We first consider eq. (\ref{twothirtyone})
for $k=0$. In this limit the average potential $U_k$ is equal
to the effective potential $U$ and the inverse propagator $P(q^2)$
assumes its standard form $P(q^2) = q^2$.
We thus obtain the well known
result for the effective potential in the leading order of the
$1/N$ expansion \cite{largen1}
\be
U'(\rho) = -\mb + \lb \rho + \frac{N}{2} \lb
\int_{\Lambda} \frac{d^4 q}{(2 \pi)^4} \frac{1}{ q^2 + U'(\rho)},
\label{throone} \ee
where we remind the reader that
$\rho = \frac{1}{2} \phi^a \phi_a$ and the derivatives of
$U$ are taken with respect to $\rho$.
The generalization of the above expression
for non-zero temperature follows (\ref{thrfffive}) and reads
\be
U'(\rho,T) = -\mb + \lb \rho + \frac{N}{2} \lb
T \sum_m \int_{\Lambda} \frac{d^{3} \vec{q}}{(2 \pi)^{3}}
\frac{1}{ \vec{q}^2 + 4 \pi^2 m^2 T^2 + U'(\rho,T)}.
\label{thrfive} \ee
\par
It is convenient to define the
functions
\beq
I(T) = &T \sum_m
\int_{\Lambda} \frac{d^3 \vec{q}}{(2 \pi)^3}
\frac{1}{\vec{q}^2+4 \pi^2 m^2 T^2}
\label{thrseven} \\
I_n(w,T) = &T \sum_m
\int_{\Lambda} \frac{d^3 \vec{q}}{(2 \pi)^3}
\frac{1}{(\vec{q}^2+4 \pi^2 m^2 T^2 + w)^n}.
\label{threight} \eeq
Then eq. (\ref{thrfive}) can be written as
\be
U'(\rho,T) = \left[ \frac{N}{2} \lb I(T)-\mb \right]
+ \lb \rho + \frac{N}{2} \lb
\left[ I_1(U'(\rho,T),T) - I(T) \right].
\label{thrnine} \ee
We are interested in solving the above equation
near the phase transition. This correponds to a temperature region in
which typically $U'(\rho,T) \ll T^2$ near the minimum of the potential.
We can, therefore, use a high temperature expansion in terms of $\frac{w}{T^2}$
for the evaluation of $I_n(w,T)$ in this regime.
In the opposite limit of zero temperature the discrete summation
can be replaced by an integration over a continuous $q^0$ (see
(\ref{thrthree})) and the integrals become four-dimensional.
Integrals similar to $I(T), I_n(w,T)$ have been discussed extensively in
the literature \cite{doljak, kapusta}. We briefly review here the
results which are relevant for our investigation.
$I(T)$ is given by
\be
I(T) = \frac{\Lambda^2}{16 \pi^2} + \frac{T^2}{12}.
\label{thrten} \ee
The first term
corresponds to the quadratic divergence of the zero temperature theory,
while for the evaluation of the second term we have made use of
$\Lambda \gg T$.
Notice that $I(T)$ receives contributions from
all values of $m$ in the infinite sum. In this sense
it incorporates effects of the effective three-dimensional theory
(modes with $m=0$) as well as the four-dimensional one ($m \not= 0$).
For the difference $I_1(w,T)-I(T)$ we find \cite{doljak}
\be
I_1(w,T)-I(T) = - \frac{1}{4 \pi} T \sqrt{w}
- \frac{w}{16 \pi^2} \ln \left( \frac{\Lambda^2}{(\ct T)^2 + w} \right),
\label{threleven} \ee
with $\ct =\exp(\frac{1}{2} + \ln(4\pi) - \gamma) \simeq 11.6$.
For $w \ll T^2$ the above expression gives the first two terms of the
high temperature expansion.
The leading term in the high temperature expansion $\sim T \sqrt{w}$
comes entirely from the $m=0$ term in the infinite sum.
As a result eq. (\ref{threleven}) in the leading order for small
$\frac{w}{T^2}$ reflects only
the dynamics of the effective three-dimensional
theory.
In contrast, the logarithmic term is typical for the four-dimensional
renormalization effects from the momentum range between
$\ct T$ and $\Lambda$.
Notice that we have included a $w$-dependence in the
argument of the logarithm which
is negligible for small $\frac{w}{T^2}$.
As a result
(\ref{threleven}) is also exact for $T=0$.
(The first term in (\ref{threleven})
is not present for large but finite $\frac{w}{T^2}$,
but since this term is small in this region
it does not affect our discussion.)
Eq. (\ref{threleven}) is therefore valid both for the high and low temperature
regime. The corrections to (\ref{threleven}) only affect an intermediate
``threshold'' region where $w \simeq (\ct T)^2$.
The integrals $I_n(w,T)$ for $n > 1$ can be
obtained by differentiation of eq. (\ref{threleven})
with respect to $w$. For later use we give
the next two integrals
\beq
I_2(w,T) =  &\frac{1}{8 \pi} \frac{T}{\sqrt{w}}
+ \frac{1}{16 \pi^2}
\ln \left( \frac{\Lambda^2}{(\ct T)^2 + w} \right)
- \frac{1}{16 \pi^2}
\frac{w}{(\ct T)^2 + w}
\label{thrtwelve} \\
I_3(w,T) =  &\frac{1}{32 \pi} \frac{T}{w^{\frac{3}{2}}}
+ \frac{1}{32 \pi^2}
\frac{2 (\ct T)^2 + w}{\left( (\ct T)^2 + w \right)^2}.
\label{thrthirteen} \eeq
Again the first term in the above expressions is determined by the
effective three-dimensional theory.
\par
Combining eqs. (\ref{thrnine}), (\ref{thrten}) and keeping only the
leading contribution in (\ref{threleven})
we obtain for $T^2 \gg U'(\rho,T)$
\be
U'(\rho,T) =
N \lb \frac{T^2}{24}
- \left[ \mb - N \lb \frac{\Lambda^2}{32 \pi^2} \right]
+ \lb \rho - N \lb \frac{T}{8 \pi} \sqrt{U'(\rho,T)}.
\label{thrfourteen} \ee
The ultraviolet divergences
can be absorbed in the $(T=0)$ renormalization of
the mass parameter and coupling.
We define the renormalized parameters
\footnote{Note that these parameters correspond to $T=0$
and do not coincide with the quadratic and quartic term in the
temperature dependent effective potential.}
for the zero temperature four-dimensional
theory \cite{largen2}
\beq
\frac{\mu^2_R}{\lambda_R} = &\frac{\mb}{\lb} - N \frac{\Lambda^2}{32 \pi^2},
\label{extraone} \\
\lambda_R = &\frac{\lb}{1 + \frac{N \lb}{32 \pi^2} \ln \left(
\frac{\Lambda^2}{M^2} \right) },
\label{extratwo} \eeq
where $M$ is an appropriate renormalization scale $M \ll \Lambda$.
Eq. (\ref{thrfourteen}) now reads
\be
U'(\rho,T) = \lb \left[ N \frac{T^2}{24} - \frac{\mu^2_R}{\lambda_R} \right]
+ \lb \rho - N \lb \frac{T}{8 \pi} \sqrt{U'(\rho,T)}.
\label{thrsixteen} \ee
The critical temperature $T_{cr}$ is defined
as the temperature at which $U'(0,\tcr)=0$.
This gives the well known result
\cite{linde2}-\cite{weinberg}
\be
T^2_{cr} = \frac{24}{N} \frac{\mu^2_R}{\lambda_R},
\label{thrfifteen} \ee
We concentrate on temperatures $T \sim \tcr$ and $\rho \ll T^2$, for which
we obtain from eq. (\ref{thrsixteen}) in leading order
\be
U'(\rho,T) = \left( \frac{\pi}{3} \frac{T^2-\tcr^2}{T} +
\frac{8 \pi}{ N} \frac{\rho}{T} \right)^2.
\label{thrseventeen} \ee
This result justifies a posteriori
the high temperature expansion for the integrals
$I_n$. Notice how any reference to the bare parameters has disappeared in
the above expression.
Simple integration gives for $U(\rho,T)$
\be
U(\rho,T) = \frac{\pi^2}{9} \left( \frac{T^2-\tcr^2}{T} \right)^2 \rho
+ \frac{1}{N} \frac{8 \pi^2}{3} \frac{T^2-\tcr^2}{T^2} \rho^2
+ \frac{1}{N^2} \frac{64 \pi^2}{3} \frac{1}{T^2} \rho^3.
\label{threighteen} \ee
The last equation
gives the effective potential in the critical region $T \sim \tcr$,
for large $N$ and including the $\rho^3$ ($\phi^6$) term.
Eq. (\ref{threighteen})
is in agreement with \cite{jain} in the leading order of the
$1/N$ expansion. It is also completely analogous to results obtained in the
context of three-dimensional field theory \cite{largen1}.
We observe an $N$ dependence for the renormalized couplings
in agreement with the general remarks of section 2 (See eq. (\ref{twotwelve}).)
Notice that the only surviving reference to the parameters of the
four-dimensional theory
exists in the definition of the critical temperature (\ref{thrfifteen}).
According to our discussion of the integrals
$I, I_n$, the critical temperature is determined by the dynamics of both the
four-dimensional and three-dimensional theory. On the contrary, the
form of the potential in the critical region
(and therefore the critical behaviour) reflects only the
effective three-dimensional theory.
\par
Eq. (\ref{threighteen}) predicts a {\bf second order} phase transition.
For $T \geq \tcr$ the theory is in the symmetric phase with the minimum of the
potential at
$\rho_0 = 0$. As the critical temperature is approached from above
the mass term is given by
\be
m_R(T) \equiv U'(0,T) = \frac{\pi}{3} \frac{T^2-\tcr^2}{T},
\label{thrnineteen} \ee
a result first obtained in \cite{doljak}.
In the language of critical exponents, the behaviour of $m_R$
corresponds to an exponent
\be
\nu = \lim_{T \rightarrow \tcr}
\frac{d \left[ \ln m_R(T) \right] }{d \left[ \ln(T-\tcr) \right] } = 1.
\label{thrtwenty} \ee
This result is in agreement with calculations of $\nu$ in the
context of the three-dimensional
zero temperature theory \cite{largen1}.
The quartic coupling in the symmetric phase
is given by
\be
\lambda_R(T) \equiv U''(0,T)
= \frac{1}{N} \frac{16 \pi^2}{3} \frac{T^2-\tcr^2}{T^2}.
\label{thrtwentyone} \ee
It goes to zero as the critical temperature is approached
with an exponent
\be
\zeta = \lim_{T \rightarrow \tcr} \frac{d \left[ \ln \lambda_R(T)
\right] }{d \left[ \ln(T-\tcr) \right] }
= 1.
\label{thrtwentytwo} \ee
The dimensionless ratio $\frac{\lambda_R(T) T}{m_R(T)}$
approaches a temperature independent value at $\tcr$
\be
\lim_{T \rightarrow \tcr}
\frac{\lambda_R(T) T}{m_R(T)} = \frac{16 \pi}{N},
\label{thrtwentythree} \ee
a result which has an analogue in the language of three-dimensional
field theory. (For an extensive discussion of this point we refer the
reader to \cite{transition,largen1,parisi}.) It is interesting to recall at
this point
that the ratio $\frac{\lb T}{m_R}$ diverges near $\tcr$ and this
is the reason for the breakdown of the standard perturbative
calculations. We see how this divergence disappears when the
bare quartic coupling is replaced by the renormalized one.
At $\tcr$ the $\rho^3$ ($\phi^6$) term is the first non-zero one,
with
a coefficient given in (\ref{threighteen}).
For $T < \tcr$ the potential has a minimum at a non-zero value
\be
\rho_0(T) = \frac{N}{24} (\tcr^2-T^2),
\label{thrtwentyfour} \ee
and assumes the form
\be
U(\rho,T) = \frac{1}{N^2} \frac{64 \pi^2}{3} \frac{1}{T^2}
\left( \rho - \rho_0(T) \right)^3,
\label{thrtwentyfive} \ee
after the addition of a $\rho$-independent constant.
Notice that the minimum of the potential (the order parameter for the
phase transition)
approaches continuously zero as the critical temperature is
approached from below (justifying our conclusion
for a phase transition of the second kind).
Expression (\ref{thrtwentyfour}) predicts a critical
exponent
\be
\beta = \lim_{T \rightarrow \tcr}
\frac{d \left[ \ln \sqrt{\rho_0(T)} \right] }{d \left[ \ln(\tcr-T) \right] }
= 0.5
\label{thrtwentysix} \ee
again in agreement with the large $N$ results of three-dimensional
field theory.
Another interesting point is the zero value of the second $\rho$-derivative
of the potential at the minimum for $T< \tcr$. This reflects the fact that
the $\phi^4$ theory is infrared free in the spontaneously broken phase
due to the Goldstone modes.
At non-zero temperature the dependence of the running coupling on the
infrared cutoff $k$ is linear \cite{transition}
(compared to the logarithmic running at zero
temperature). This results in a zero renormalized coupling
$\lambda_R(T) = U''(\rho_0,T)$
in the limit $k \rightarrow 0$.
We see how this physical effect is naturally incorporated in
our result (\ref{thrtwentyfive}).
\par
A remark is due at this point about the form of the potential
(\ref{thrtwentyfive}), since, obviously, the requirement of
convexity is not satisfied.
The paradox is resolved if one realizes
that
our discussion, and therefore the result (\ref{thrtwentyfive}),
is valid only in the
``outer'' region of the potential $\rho \geq \rho_0$.
For $\rho < \rho_0$ the true saddle point
of the functional integral (\ref{twonine}), (\ref{twoten}) corresponds to
$\cb = 0$ and $\bb$ given by the minimum of $V(B)$.
This saddle point is closely related to the spin wave solution
of ref. \cite{christof, convex}. It results in
the convex envelope of the non-convex part of
(\ref{thrtwentyfive}).
\par
The calculation of terms with higher powers of $\rho$ in the effective
potential
requires additional terms in the high temperature expansion.
Inserting (\ref{thrten}) and (\ref{threleven}) into (\ref{thrnine}) and
using (\ref{extraone}), (\ref{extratwo}) for the definition of
$\mu_R^2$ and $\lambda_R$, one obtains
\be
U'(\rho,T) \left[ \frac{1}{\lambda_R} + \frac{N}{32 \pi^2}
\ln \left( \frac{M^2}{(\ct T)^2 + U'(\rho,T)} \right)
\right]
=
\left[ N \frac{T^2}{24}
-  \frac{\mu^2_R}{\lambda_R}
\right]
+ \rho - N \frac{T}{8 \pi} \sqrt{U'(\rho,T)}.
\label{thrtwentyseven} \ee
Differentiation with respect to $\rho$ gives the $\rho-$ and $T$-dependent
quartic coupling
\be
U''(\rho,T) =
\frac{\lambda_R}
{
1 + \frac{N \lambda_R}{32 \pi^2}
\left[
\ln \left( \frac{M^2}{(\ct T)^2 + U'(\rho,T)} \right)
- \frac{U'(\rho,T)}{(\ct T)^2 + U'(\rho,T)}
+ \frac{2 \pi T}{\sqrt{U'(\rho,T)}}
\right]
}.
\label{telosone} \ee
It is interesting to study its dependence on the effective infrared
cutoff $U'(\rho,T)$. With $\tilde{t}= \frac{1}{2}\ln U'$ one finds
\be
\frac{\partial}{\partial \tilde{t}} U''
= \frac{N}{16 \pi^2} (U'')^2 \left[
\frac{U'(U'+2(\ct T)^2)}{(U'+(\ct T)^2)^2}
+ \pi \frac{T}{\sqrt{U'}} \right].
\label{telostwo} \ee
We observe the transition from the usual logarithmic four-dimensional
running
\footnote{Note that, for $U' \gg (\ct T)^2, |\mu_R^2|$, one has
approximately $U' \simeq \lambda_R \rho$ and (\ref{telostwo})
reproduces the result of \cite{colwein} in leading order.}
for $U' \gg (\ct T)^2$ to the effective linear three-dimensional
running \cite{transition,oconnor} for $U' \ll (\ct T)^2$.
\par
The critical temperature is again given by
(\ref{thrfifteen}).
For $T$ near $\tcr$ and small $\rho$ we can solve
(\ref{thrtwentyseven}) approximately. Including the next to leading term
we find, with $M = \ct T$,
\be
U(\rho,T) =
\frac{1}{N^2} \frac{64 \pi^2}{3} \frac{1}{T^2}
\left( \rho +
\frac{N}{24} (T^2 - \tcr^2)
\right)^3
-
\frac{1}{N^3} \frac{1}{\lambda_R N} 2048 \pi^4 \frac{1}{T^4}
\left( \rho +
\frac{N}{24} (T^2 - \tcr^2)
\right)^4,
\label{thrtwentynine} \ee
with higher order terms involving larger powers of
$\frac{8 \pi^2}{N \lambda_R T^2} \left( \frac{T^2-\tcr^2}{3}
+ \frac{8 \rho}{N} \right)$.
The second term
in (\ref{thrtwentynine}) introduces
corrections
to the
the effective potential
(\ref{threighteen})
which are not universal with respect to the three-dimensional effective
theory. These corrections are sensitive to the properties of the full
four-dimensional theory: they depend on $\lambda_R$. We note that the
range where the pure three-dimensional behaviour dominates
\be
|T^2 - \tcr^2 + \frac{24}{N} \rho | \ll \frac{3 N \lambda_R T^2}{ 8 \pi^2}
\label{ttelos} \ee
is proportional to $\lambda_R$ and, thus, very small for small $\lambda_R$.
In the symmetric phase ($T \geq \tcr$) and in the critical region
$T \rightarrow \tcr$,
the corrections to the first three terms in the
$\rho$ expansion are suppressed
by additional powers of $\frac{T^2-\tcr^2}{T^2}$. Keeping the leading
contibution we obtain
for the $\rho^4$ ($\phi^8$) coupling
\be
U^{(4)}(0,T) =
- \frac{49152 \pi^4}{N^4 \lambda_R T^4}.
\label{thrthirty} \ee
The negative sign of the $\phi^8$ term does not imply
the presence of a second minimum. It is obvious from
eq. (\ref{thrtwentyseven}) that above $\tcr$ there is only one
minimum at $\rho=0$.
For $T< \tcr$ the minimum of the effective potential is
given by (\ref{thrtwentyfour}) and the potential assumes the form
\be
U(\rho,T) = \frac{1}{N^2} \frac{64 \pi^2}{3} \frac{1}{T^2}
\left( \rho - \rho_0(T) \right)^3
-
\frac{1}{N^3} \frac{1}{\lambda_R N} 2048 \pi^4 \frac{1}{T^4}
\left( \rho - \rho_0(T) \right)^4.
\label{thrthirtyone} \ee
Eqs. (\ref{thrtwentynine}), (\ref{thrthirtyone}) can be directly compared to
(\ref{threighteen}), (\ref{thrtwentyfive}). We conclude that the universal
behaviour near the critical temperature
(which determines the nature of the transition and the critical behaviour of
the system) is
effectively three-dimensional
and is given by the first three derivatives
of the potential around its minimum (the ``relevant'' terms).
Higher derivatives of the potential do not affect the universal behaviour.
They simply involve secondary
non-universal corrections to the effective potential.
One should keep in mind, however, that the non-universal corrections
(which reflect the four-dimensional nature of the theory)
must be taken into account for temperatures away from the critical one.
\par
In summary, the universal part of the effective potential for the $\phi^4$
theory in the critical region
is given by eq. (\ref{threighteen})
for large $N$.
In the following section, we shall reproduce this result by solving the
Schwinger-Dyson equations for the high temperature theory
near the phase transition.

\setcounter{equation}{0}
\renewcommand{\theequation}{{\bf 4.}\arabic{equation}}

\section*{4. The Schwinger-Dyson equations}

In this section we are interested in verifying the results of
our previous discussion through a method more closely
connected to the standard perturbative evaluation of the
effective potential \cite{colwein,linde2,weinberg,doljak}.
It is well known \cite{doljak}
that the Schwinger-Dyson equation for the propagator,
for a $\phi^4$ theory in the large $N$ limit, can be reproduced
by the summation of an infinite class of diagrams, usually characterized as
``daisy'' or ``cactus'' graphs \cite{doljak,linde2}.
The solution of this equation gives the value of the critical temperature
and the temperature dependence of the mass term in the critical
region.
We have seen in the previous section that the
first three terms of the effective potential
in the critical region (\ref{threighteen}) do not involve the
parameters of the
four-dimensional theory, but are fully determined by the dynamics of
the effective three-dimensional theory.
Moreover, the form of the potential predicts a characteristic temperature
dependence for its second and third $\rho$-derivatives (which correspond to the
renormalized four and six point vertices).
These results have so far no analogue in the context of perturbation
theory at non-zero temperature,
since the renormalization of the quartic coupling has not yet been considered
\cite{wrong,zwirner}.
It is interesting to investigate whether the behaviour predicted by eq.
(\ref{threighteen})
can be reproduced through the study of the relevant Schwinger-Dyson equations,
since a solution of these equations automatically accounts for the
renormalization of the couplings.
\par
In fig. 1 we give a graphical representation
of the first three
Schwinger-Dyson equations for the $O(N)$ $\phi^4$ theory in the symmetric
phase for large $N$.
(For
a derivation of the
Schwinger-Dyson equations in the functional integral formalism see
\cite{itzykson}.)
Only the leading terms in $1/N$ have been included. (We remind the reader
that the
vertices come with an power of $N$ given by (\ref{twotwelve}).)
The left-hand-side of the equations involves the first three $\rho$-derivatives
of the potential at zero external momentum (the renormalized mass and
four and six point vertices).
Simple vertices denote bare quantities, and lines
with a circle full propagators.
The representation is schematic and the signs and combinatoric factors for the
various terms have not been included.
The incorporation of
non-zero temperature effects in the formalism proceeds exactly as in the
previous section.
By recalling eqs. (\ref{thrseven}), (\ref{threight}) we can express eq. (a)
of fig. 1 as
\be
U'(0,T) = \left[ \frac{N}{2} \lb I(T)-\mb \right]
+ \frac{N}{2} \lb
\left[ I_1(U'(0,T),T) - I(T) \right].
\label{fourone} \ee
For the solution of the above equation we proceed as for the solution
of (\ref{thrnine}):
We use eqs. (\ref{thrten}) and (\ref{threleven}),
introduce the renormalized quantities
$\mu^2_R, \lambda_R$
according to (\ref{extraone}), (\ref{extratwo}),
define the critical temperature (\ref{thrfifteen}), and finally
obtain the standard result \cite{doljak}
\be
U'(0,T) = m^2_R(T) = \left[ \frac{\pi}{3} \frac{T^2-\tcr^2}{T} \right]^2.
\label{fourtwo} \ee
Eq. (b) of fig. 1 reads in our notation
\be
U''(0,T) = \lb - \frac{N}{2} \lb U''(0,T)~ I_2(U'(0,T),T),
\label{fourthree} \ee
where $I_2$ was defined according to (\ref{threight}).
Making use of the high temperature expansion (\ref{thrtwelve}) we
find
\be
U''(0,T) = \frac{\lb}{1 + \frac{N}{16 \pi} \lb T [U'(0,T)]^{-\frac{1}{2}}}.
\label{fourfour} \ee
In the critical region
$T \rightarrow \tcr$ the second term in the denominator of the last expression
(which comes entirely from the effective three-dimensional theory) diverges.
Substitution of (\ref{fourtwo}) gives
\be
U''(0,T) = \lambda_R(T) = \frac{1}{N} \frac{16 \pi^2}{3}
\frac{T^2-\tcr^2}{T^2}.
\label{fourfive} \ee
Notice that all reference to $\lb$ has again disappeared in the
above equation.
In a similar way we obtain for eq. (c) of fig. 1
\be
U'''(0,T) = - \frac{N}{2} \lb U'''(0,T)~ I_2(U'(0,T),T)
+ N \lb \left[ U''(0,T) \right]^2 I_3(U'(0,T),T).
\label{foursix} \ee
Through use of (\ref{thrtwelve}), (\ref{thrthirteen}), (\ref{fourtwo}),
(\ref{fourfive}) we find
\be
U'''(0,T) = \frac{1}{N^2} 128 \pi^2 \frac{1}{T^2}.
\label{fourseven} \ee
\par
Reconstruction of the effective potential in the symmetric phase,
through a Taylor expansion around the origin, results
in eq. (\ref{threighteen}). It should also be noted that eqs.
(\ref{fourone}), (\ref{fourthree}), (\ref{foursix}) can be obtained
from (\ref{thrfive}) by differentiation with respect to $\rho$.
We have, therefore, shown that the solution of the
Schwinger-Dyson equations in the large $N$ limit
verifies the results of the discussion
of the previous section. Moreover, it provides helpful insight on
how the
first three terms of the
effective potential for $T \rightarrow \tcr$ become
independent of the four-dimensional parameters, due to
the singular infrared behaviour of the effective three-dimensional theory.
In the following section we shall use the evolution equation for the
effective potential in order to derive results
beyond the leading order in $1/N$.

\setcounter{equation}{0}
\renewcommand{\theequation}{{\bf 5.}\arabic{equation}}

\section*{5. Beyond leading order in $1/N$}

The picture for the high temperature phase transition for the
$O(N)$-symmetric scalar theory is completely satisfactory for large
values of $N$.
In order to deal with realistic theories, e.g. $N=4$ for the scalar
sector of the electroweak standard model or the QCD phase transition
for vanishing up and down quark mass \cite{wilczek},
one has to go beyond the leading order approximation and include corrections
$\sim 1/N$. Near the phase transition the three-dimensional behaviour is
dominant and any useful method should account for a proper treatment of the
three-dimensional $O(N)$-symmetric scalar theory. In particular,
one expects the critical exponents $\nu, \beta, \zeta$ to deviate from their
simple values
(\ref{thrtwenty}), (\ref{thrtwentysix}), (\ref{thrtwentytwo}), so that the
effective potential becomes non-analytic in $T-\tcr$.
\par
Going beyond the leading result in $1/N$ is
not as straightforward as it may seem at first sight.
The difficulty lies in the infrared behaviour of the three-dimensional
theory. At the critical temperature, this leads to the infrared
divergencies of standard high temperature perturbation theory.
A systematic expansion of (\ref{twonine}), (\ref{twoten}) beyond the
saddle point approximation is not easy, even for
$k=0, P(- \Box) = - \Box$ and $\phi^a =$ const.
As far as the universal behaviour at the critical temperature is concerned, one
can rely, of course, on results of the purely three-dimensional $1/N$ expansion
\cite{largen1}. This should correctly predict the first $1/N$ correction
to the critical exponents. However, for an estimate of either the critical
temperature, or the behaviour of $U(\rho,T)$ away from the critical
temperature,
or the proportionality constants in relations such as $m_R \sim
(T-\tcr)^{\nu}$,
the properties of the underlying four-dimensional theory play a role.
For these questions the four-dimensional $1/N$ expansion at
non-vanishing temperature is needed. The same remark applies
to other standard methods employed in three-dimensional field theory such
as the $4-\epsilon$ or $2+\epsilon$ expansions or the strong coupling
results on the lattice. They give reliable results for the critical behaviour
but cannot be employed for the behaviour
away from $\tcr$, where the matching with the four-dimensional behaviour is
needed.
\par
One could, in principle, consider the Schwinger-Dyson equations and
include terms subleading in $1/N$. These terms, however, turn
the Schwinger-Dyson equations into a complicated coupled system, as illustrated
in fig. 2.
The computation of the two point function involves now the full four point
vertex and this difficulty propagates to all vertices. As we have
seen before, a replacement of the full four point vertex by the bare
coupling $\lb$ does not reflect properly the infrared behaviour near
$\tcr$, where the appropriate renormalized coupling $\lambda_R(T)$ vanishes.
More precisely, the last diagram in fig. 2 with $\lb$ instead of the full
four point vertex is infrared divergent at $T=\tcr$, whereas the correct
expression is infrared finite.
\par
All these effects are naturally incorporated in the evolution equation
(\ref{twotwentyseven}), which has a simple generalization
\cite{christof,transition} for arbitrary $N$. When wavefunction
renormalization is neglected, the generalized evolution equation reads
\footnote{
This equation can also be interpreted as a truncation of an exact
evolution equation \cite{exact} (for arbitrary $N$) which is related to
an improved average action \cite{more}.}
\beq
k \frac{d}{dk} U_k(\rho) = &
\frac{1}{2}
\int_{\Lambda} \frac{d^dq}{(2 \pi)^d}
\left(
\frac{N-1}{ P(q^2) + U'_k(\rho)} +
\frac{1}{ P(q^2) + U'_k(\rho) + 2 U''_k(\rho) \rho}
\right)
k \frac{d}{dk} P(q^2)  \nonumber \\
= &
v_d
\int_{0}^{\Lambda^2} {dx} x^{\frac{d}{2}-1}
\left(
\frac{N-1}{ P(x) + U'_k(\rho)} +
\frac{1}{ P(x) + U'_k(\rho) + 2 U''_k(\rho) \rho}
\right)
k \frac{d}{dk} P(x),
\label{fivone} \eeq
where $v_d$ is given by (\ref{twotwentyfive}).
One can easily identify in the above expression the contributions
of the $N-1$ Goldstone modes and the radial mode. In the limit
$N \rightarrow \infty$ the contribution of the
Goldstone modes dominates and the evolution
equation becomes identical to (\ref{twotwentyseven}).
The generalization of the last equation in order to include non-zero
temperature effects is straightforward and follows the discussion
in the end of section 2.
For the four-dimensional theory one finds
\newpage
\beq
k \frac{d}{dk} U_k(\rho,T) = &
\frac{1}{2} T \sum_m
\int_{\Lambda} \frac{d^3 \vec{q}}{(2 \pi)^3}
k \frac{d}{dk} P(\vec{q}^2 + 4 \pi^2 m^2 T^2) \nonumber \\
& \cdot \left(
\frac{N-1}{ P(\vec{q}^2 + 4 \pi^2 m^2 T^2) + U'_k(\rho,T)} +
\frac{1}{ P(\vec{q}^2 + 4 \pi^2 m^2 T^2) + U'_k(\rho,T)+ 2 U''_k(\rho,T) \rho}
\right) \nonumber \\
{}~~&~~
\label{fivtwo} \eeq
For $k \gg T$ the summation
over the discrete values of $m$ in the above equation is equal to an
integration over a continuous range of $q^0$, up to exponentially small
corrections. In this limit eq. (\ref{fivtwo}) is identical to (\ref{fivone})
with $d=4$. In the opposite limit $k \ll T$
the summation over $m$ is dominated by the $m=0$ contribution and the
evolution becomes effectively three-dimensional. The full solution of
(\ref{fivtwo}) interpolates between the two regimes
and, thus, matches the four-dimensional behaviour
with the three-dimensional one.
\par
For the study of the phase transition we are interested in solving
(\ref{fivtwo}) near the minimum of the potential for different
temperatures $T$.
In order to
do so we parametrize $U_k(\rho,T)$ in terms of the location
of its minimum $\rho_0(k,T)$ and
its successive $\rho$-derivatives at the minimum
$U_k'(\rho_0,T), U''_k(\rho_0,T) ...$ This turns the
partial differential equation (\ref{fivtwo}) into an infinite system of
coupled differential equations, which can be solved approximately
by truncation at
a finite number of equations. We have seen in the previous sections that
the order of the phase transition as well as the critical behaviour of the
system are determined by the
``relevant'' first three $\rho$-derivatives of the effective
potential. It is sufficient, therefore, to keep the first three coupled
differential equations of the infinite system.
We give these equations
without presenting  the details of their derivation
here, since they have been repeatedly
derived elsewhere \cite{christof,transition,
convex}. In the \underline{spontaneously broken regime}
($\rho_0 \not= 0, U'_k = 0$)
the equations for $\rho_0(k,T)$,
$U''_k(\rho_0,T)$ and
$U'''_k(\rho_0,T)$
read
\beq
\frac{d \rrz}{dt} =
&-v_4 k^{2} \biggl\{ \left( 3 + \frac{2 U'''_k \rrz}{U''_k} \right)
L^4_1(2 U''_k \rrz,T) + (N-1) L^4_1(0,T) \biggr\}
\label{fivthra} \\
\frac{d U''_k}{dt} =
&-v_4 [U''_k]^2 \biggl\{ \left( 3 + \frac{2 U'''_k \rrz}{U''_k} \right)^2
L^4_2(2 U''_k \rrz,T) +
(N-1) L^4_2(0,T) \biggr\} \nonumber \\
&+ v_4 k^{2} \left( 2 U'''_k - \frac{2 [U'''_k]^2 \rrz}{U''_k} \right)
L^4_1(2 U''_k \rrz,T)
\label{fivthrb} \\
\frac{d U'''_k}{dt} =
&~~2 v_4 k^{-2}
[U''_k]^3 \biggl\{ \left( 3 + \frac{2 U'''_k \rrz}{U''_k} \right)^3
L^4_3(2 U''_k \rrz,T) +
(N-1) L^4_3(0,T) \biggr\} \nonumber \\
&- 3 v_4 U''_k U'''_k
\biggl\{ 5 \left( 3 + \frac{2 U'''_k \rrz}{U''_k} \right)
L^4_2(2 U''_k \rrz,T) +
(N-1) L^4_2(0,T) \biggr\}.
\label{fivthrc}
\eeq
In the \underline{symmetric regime} ($\rho_0 = 0$)
the corresponding equations for
$U'_k(0,T)$,
$U''_k(0,T)$ and
$U'''_k(0,T)$
read
\beq
\frac{d U'_k}{d t} = &
{}~~v_4 k^{2} (N+2) U''_k L^4_1(U'_k,T)
\label{fivfoura} \\
\frac{d U''_k}{d t} = &
- v_4 (N+8) [U''_k]^2 L^4_2(U'_k,T)
+ v_4 k^2 (N+4) U'''_k L^4_1(U'_k,T)
\label{fivfourb} \\
\frac{d U'''_k}{d t} = &
{}~~2 v_4 k^{-2} (N+26) [U''_k]^3 L^4_3(U'_k,T)
- 3 v_4 (N+14) U''_k U'''_k L^4_2(U'_k,T).
\label{fivfourc}
\eeq
The dimensionless integrals $L^4_n(w,T)$
\be
L^4_n(w,T) =
- n k^{2n-4}
\frac{2}{\pi}
 T \sum_m
\int_{\Lambda} d^3 \vec{q}
\left( P(\vec{q}^2 + 4 \pi^2 m^2 T^2) + w \right)^{-(n+1)}
k \frac{d}{dk} P(\vec{q}^2 + 4 \pi^2 m^2 T^2)
\label{fivfive} \ee
have been extensively discussed in \cite{transition}.
(Notice also that these integrals are proportional to the
logarithmic derivative with respect to $k$ of the integrals
$I_n$ discussed in section 3, with $q^2$ replaced by $P(q^2)$ in the
propagator.)
They exhibit a behaviour as a function of
$\frac{T}{k}$ which interpolates between being
purely four-dimensional for $T \ll k$ and
effectively three-dimensional for $T \gg k$.
This induces
a continuous transition from four to three for
the effective dimensionality
of the evolution equations.
\par
The high temperature
phase transition for the $O(N)$-symmetric $\phi^4$ theory
was investigated in detail in ref. \cite{transition}.
We shall not repeat the technical discussion here and we refer the reader to
\cite{transition} for a thorough presentation.
We briefly summarize the main results in order to
point out the consistency with the conclusions of the previous sections.
The evolution equation was studied through the parametrization of the
average potential in terms of its minimum and its derivatives at the minimum,
as explained earlier in this section. The resulting infinite system of
differential equations was truncated after two equations and
only the first two derivatives of the
potential were considered. This amounts to solving the equations
(\ref{fivthra}), (\ref{fivthrb}) and (\ref{fivfoura}), (\ref{fivfourb})
without the terms involving $U'''_k$.
The evolution equations were solved at various temperatures
in order to obtain the
parameters of the theory in the limit $k \rightarrow 0$.
In this limit the average potential becomes equal to the
effective
potential, and its minimum gives the ground state of the
theory and its derivatives the renormalized mass and couplings.
The phase transition was studied by
defining the renormalized theory in the spontaneously broken
phase at $T=0$, and following the evolution of
the minimum of the effective potential
for increasing temperature. The phase transition was found to be of the
second order
with a critical temperature
given by (\ref{thrfifteen})
for small values of the quartic
coupling, independent of $N$.
The theory was shown to
have a vanishing quartic coupling at and below
the critical temperature
for any
$N \not= 1$, in agreement with (\ref{thrtwentyfive}).
The universal critical behaviour was found to be determined by
a fixed point corresponding to a second order phase transition.
The same fixed point determines
the critical behaviour of the effective three-dimensional theory, in
exact agreement
with the discussion in sections 3 and 4.
An extrapolation
of the numerical results for the critical exponents
in \cite{transition},
to large values of $N$,
indicates
agreement with (\ref{thrtwenty}), (\ref{thrtwentytwo}),
(\ref{thrtwentysix}). Evaluation of the analytic expressions
for the exponents
in \cite{transition} for $N \rightarrow \infty$
reproduces the same values.
Finally the numerical values for the ratio
$\frac{\lambda_R (\tcr) \tcr}{m_R(\tcr)}$
are consistent with (\ref{thrtwentythree})
for large values
of $N$.
\par
We have repeated the calculation of the critical exponents by
considering the first three derivatives of the potential. This was done by
solving the system of evolution equations
(\ref{fivthra})-(\ref{fivthrc}) and (\ref{fivfoura})-(\ref{fivfourc}).
In this way all the
parameters
``relevant'' for the universal critical behaviour (apart from the
wavefunction renormalization which is non-zero for finite $N$)
were taken into account. The results of this calculation for
the exponent $\beta$ (for a value $b=3$, see (\ref{twofour}))
are presented in table 1. The results for the other exponents
satisfy
$\nu = 2 \beta$ (in agreement with the scaling law for zero
wavefunction renormalization) and $\zeta=\nu$.
In the same table we list values for the exponent
$\beta$ obtained through the
$1/N$ expansion to order $1/N^2$
for the three-dimensional theory \cite{journal}. We observe exact agreement for
large values of $N$, which becomes bad for small values of $N$.
For $N=3$ we also give the value for the same critical exponent obtained
through other methods ($\epsilon$-expansion,
summed perturbation theory in three
dimensions, lattice calculations) \cite{largen1}. It
is clear that our result is in good
agreement with this value, while the $1/N$
expansion fails for small $N$.
The small remaining discrepancy is due to the
omission of the wavefunction renormalization from the evolution
equations. The study of the wavefunction renormalization
effects is under way \cite{preparation}.
\par
We summarize this section by
concluding that the solution of the evolution equation fully confirms the
results of the previous two sections, while it provides the means
for the detailed study of the phase transition for small values of $N$.
We also emphasize that this method gives a reliable description
for temperatures away from the critical one \cite{transition}.

\setcounter{equation}{0}
\renewcommand{\theequation}{{\bf 6.}\arabic{equation}}

\section*{6. Conclusions}

We have studied the high temperature
phase transition for the $N$-component $\phi^4$
theory in the large $N$ limit, in order to confirm the results of
ref. \cite{transition} in which the method
of the average potential was employed.
The investigation was carried out through use of different methods, such as
the standard large $N$ evaluation of the functional integral for the
effective potential and the solution of the
Schwinger-Dyson equations.
Our results fully confirm the results of \cite{transition}.
The phase transition is of the second kind and falls in the same
universality class as the phase transition of the three-dimensional
$O(N)$-symmetric $\sigma$-models.
The critical behaviour is determined by the
first three
derivatives with respect to $\phi^2$ of the effective potential
around its minimum, and it is effectively three dimensional.
Explicit expressions have been obtained for the critical temperature
and the effective potential in the critical region.
As the critical temperature is approached from above, the
renormalized mass term $m_R(T)$ in the potential approaches zero
with a critical exponent $\nu=1$. The renormalized quartic coupling
$\lambda_R(T)$ approaches zero also, with a critical exponent $\zeta=1$.
The ratio $\frac{\lambda_R(T) T}{m_R(T)}$ remains finite and, as a result,
no infrared divergencies arise.
At the critical temperature the first non-zero term in the potential
is the $\phi^6$ term, whose coefficient has been calculated.
Below the critical temperature the minimum of the potential
(which gives the ground state and, therefore, the
order parameter of the theory)
moves continuously away from zero, with a temperature
dependence characterized by an exponent $\beta=0.5$.
All the above is fully consistent with the results of ref. \cite{transition}
and three-dimensional field theory.
\par
Going beyond the leading $1/N$ result, in order to discuss small,
physically relevant values of $N$, is more difficult and the
renormalization group becomes an indispensable tool.
We have demonstrated the power of the method of the average potential
by calculating the critical exponents for various values of $N$.
We have found perfect agreement with calculations
using the $1/N$ expansion in three dimensions. Morevoer, we have
shown that the average potential gives the correct answer even for small
values of $N$, where the $1/N$ expansion fails.
The final ingredient for the complete description of the phase
transition is the wave function renormalization, and this investigation is
under way \cite{preparation}.

\subsection*{Acknowledgements}

We would like to thank W. Buchm\"uller for discussions.

\newpage
\section*{Tables}

\begin{table} [h]
\renewcommand{\arraystretch}{1.5}
\hspace*{\fill}
\begin{tabular}{|c||c|c|}	\hline
$N$  &\multicolumn{2}{c|}{$\beta$} \\
\hline \hline
$100$ & 0.50 & $0.50^{(a)}~~~~~~~~~$
\\ \hline
$20$ & 0.48 & $0.48^{(a)}~~~~~~~~~$
\\ \hline
$10$ & 0.46 & $0.46^{(a)}~~~~~~~~~$
\\ \hline
$4$ & 0.41 & $0.34^{(a)}~~~~~~~~~$
\\ \hline
$3$ & 0.40 & $0.25^{(a)}~0.38^{(b)}$
\\ \hline
\end{tabular}
\hspace*{\fill}
\renewcommand{\arraystretch}{1}
\caption[y]
{Critical exponent $\beta$ for various $N$ ($b=3$).
For comparison we have listed
the results of calculations of the same exponent in three-dimensional
field theory:\\
a)~From $1/N$ expansion to order $1/N^2$ \cite{journal}. \\
b)~Average value of results
from $\epsilon$-expansion, summed perturbation theory in three
dimensions
and lattice calculations \cite{largen1}.
}
\end{table}

\newpage
\section*{Figures}
\vspace{140mm}

\renewcommand{\labelenumi}{Fig. \arabic{enumi}}
\begin{enumerate}
\item  
Schwinger-Dyson equations in the large $N$ limit.
\vspace{60mm}
\item  
Schwinger-Dyson equation for the two point function.
\end{enumerate}

\end{document}